\documentclass[pre,twocolumn,amsmath,amssymb,superscriptaddress,longbibliography,showpacs,nofootinbib]{revtex4-1}

\usepackage{color}
\usepackage{latexsym}

\usepackage{bm, mathtools}
\usepackage{graphicx}
\usepackage{graphics,epsfig,color}
\usepackage{dcolumn}
\usepackage{times}
\usepackage[english]{babel}
\usepackage[latin1]{inputenc}
\usepackage[T1]{fontenc}

\usepackage{cellspace}%
\usepackage{makecell}
\setcellgapes{3pt}

\newcommand{\caD}{{\cal D}}
\newcommand{\mean}[1]{\left< #1 \right>}

\newcommand{\nn}{\nonumber\\}
\newcommand{\intii}{\int_{-\infty}^{\infty}}
\newcommand{\pder}[2]{\frac{\partial#1}{\partial#2}}
\newcommand{\npder}[3]{\frac{\partial^#3 #1}{\partial#2^#3}}

\newcommand{\zpder}[2]{{\partial_{#2}#1}}

\definecolor{linkcolor}{rgb}{0,0,0.6} 
\usepackage[pdftex,colorlinks=true, pdfstartview=FitV, linkcolor= linkcolor, citecolor= linkcolor, urlcolor= linkcolor, hyperindex=true,hyperfigures=true]{hyperref} 

\begin{document}

\title{Generalized equipartition for nonlinear multiplicative Langevin dynamics:\\
  application to laser-cooled atoms}

\author{Gianmaria Falasco}
\email{gianmaria.falasco@uni.lu}
\affiliation{Complex Systems and Statistical Mechanics, Physics and Materials Science Research Unit, University of Luxembourg, L-1511 Luxembourg, Luxembourg}

\author{Eli Barkai}
\email{barkaie@biu.ac.il}
\affiliation{Department of Physics, Bar Ilan University, Ramat-Gan 52900, Israel}

\author{Marco Baiesi}
\email{baiesi@pd.infn.it}
\affiliation{
Department of Physics and Astronomy, University of Padova, 
Via Marzolo 8, I-35131 Padova, Italy
}
\affiliation{
INFN, Sezione di Padova, Via Marzolo 8, I-35131 Padova, Italy
}

\date{\today}

\begin{abstract} 
The virial theorem, and the equipartition theorem in the case of quadratic degrees of freedom, are handy constraints on the statistics of equilibrium systems. Their violation is instrumental in determining how far from equilibrium a driven system might be.
We extend the virial theorem to nonequilibrium conditions for Langevin dynamics with nonlinear friction and multiplicative noise. In particular, we generalize the equipartition theorem for confined laser-cooled atoms in the semi-classical regime. 
The resulting relation between the lowest moments of the atom position and velocity allows to measure in experiments how dissipative the cooling mechanism is. Moreover, its violation can reveal the departure from a strictly harmonic confinement or from the semi-classical regime.
 
\end{abstract}


\maketitle

\section{Introduction}

A major challenge in the field of condensed matter is to quantify how far from equilibrium a complex system is. In both hard and soft matter experiments it is often unclear how driving and dissipation---typically operating locally and microscopically---manifest in the large scale dynamics. Such complexity makes detailed measures of dissipation difficult to perform. 

Entropy production, which is proportional to the dissipated heat for a system in contact with a single thermal bath \cite{maes03,sei12}, is the most used measure of nonequilibrium \cite{cil17}. Nevertheless, it depends on all the nonequilibrated degrees of freedom of a system, thus making it problematic to be directly measured \cite{rol10}. Oftentimes, entropy production is estimated by `local' measurement. Indeed, recent advances in nonequilibrium physics have showed that entropy production can be lower bounded by the signal-to-noise ratio of a class of observables \cite{bar15a,hor20,dec20}. Although these inequalities may be useful for stationary Markovian systems, they are loose for non-stationary and non-Markovian dynamics \cite{fal20}.
Moreover, entropy production becomes useless for systems exhibiting slow dynamics and long-lived metastable states. Indeed, it can be evaluated divergent on experimental timescales \cite{mur14} because some backward transitions cannot be observed.

The probability currents in some reduced phase space can be considered instead of entropy production \cite{zia07}. Even though the method can be useful to detect broken detailed balance \cite{gne18, li19}---revealed by the presence of any statistically significant current---and may be applicable to slow systems, it suffers from some of the aforementioned downsides. Namely, a correct estimation of the dissipation cannot be achieved by monitoring only a portion of the global phase space of the system.

Another common approach consists in comparing the spontaneous fluctuations of the systems with its linear response to an external perturbation. Violation of the proportionality between the two---the tenet of the fluctuation-dissipation theorem, valid in equilibrium \cite{bai13}---is then taken as a measure of nonequilibrium \cite{cug97, har05, lip14, lip05, bai09}. Clearly, this method is not free from drawbacks either. In particular, the fact that the perturbation may inadvertently force the system into the nonlinear regime and the strong dependence on the choice of the observable make the method of limited efficacy.

The virial theorem is a further result that can reveal signatures of nonequilibrium.
Recently, it was noted that it applies to a large class of stationary states---not necessarily detailed balanced---in a generalized form, and reduces to the commonly known formula under equilibrium conditions \cite{fal16v}. The relevance of these results is twofold. On the one hand, the violation of the equilibrium virial theorem---or the equipartition theorem for the case of quadratic degrees of freedom---can be used as a measure of nonequilibrium. On the other hand, the violation of its generalized expression (valid away from equilibrium) can reveal the breakdown of the conditions under which it holds true, namely stationary Langevin dynamics with additive noise and linear friction.

Here, we further extend the result of \cite{fal16v} by deriving a generalized virial theorem and the corresponding equipartition law for Langevin dynamics with multiplicative noise and non-linear friction. We then specialize the results to the semi-classical model of trapped atoms \cite{mar96,lut13,dec15,dec16,afe21}, laser-cooled by the Sisyphus mechanism. 
The violation of the classical equipartition theorem for this system was experimentally observed by Afek et al.~\cite{afe20}, which is however a combined effect of dissipative cooling and anharmonicity of the confining lattice potential.
Our generalized virial and equipartition relations, \eqref{laser} and \eqref{laser+h}, respectively, provide a concrete way to disentangle these effects. Indeed, the presence of a nonharmonic external potentials makes \eqref{laser+h} invalid, while \eqref{laser} holds true; the presence of dissipation makes the combination of the higher order moments non-vanishing in both expressions. Besides that, the simultaneous breakdown of both formulas 
would signify a breakdown of the semi-classical Fokker-Planck description, which might happen in experiments when atoms are not trapped in a deep lattice.

\section{Virial for Langevin dynamics}

In this section we derive a nonequilibrium relation between average kinetic energy and force-dependent terms. It builds upon the mesoscopic virial equation~\cite{fal16v}, here with a special focus on nonlinear frictions. We limit our discussion to cases where position-dependent forces are conservative and derived from a confining potential $U$, while all nonequilibrium effects are explicitly appearing only in the velocity-dependent forces, which we associate to the notion of nonlinear frictions. A generalization to nonconservative position-dependent forces would be straightforward and was already exposed~\cite{fal16v}. Its derivation used (backward) generators of the dynamics while here we use the complementary approach: we start from densities and use the velocity dependent Fokker-Planck equation, also known as Kramers equation.

Let us first briefly consider a Langevin motion for a momentum $p=m\,dq/dt$, where $m$ is the mass of the system.
When particles are in an external potential field $U(q)$, a deterministic space-dependent force $-U'\equiv -\zpder{U}{q}$ enters in the Langevin equations
\begin{align}
  \dot p(t) & =  -U'(q) + F(p) + \sqrt{2 D(p)} \cdot \xi(t) \nonumber \\
  \dot q(t) & = p(t) / m
  \label{lange}
\end{align}
where $\xi(t)$ is a standard white noise and $D(p)=D_0 + D_1(p)$ is a diffusion coefficient that in general may have a term $D_1(p)$ depending  on the momentum. We choose to interpret this equation in the Ito convention for stochastic calculus and to denote this choice by the dot in $\sqrt{2 D(p)} \cdot \xi(t)$.
The term $F(p(t))$ represents a momentum-dependent force. It contains a friction force $F_{f}(p)$, in general different from the linear damping $\sim - p$ of the original Langevin equation, yet we assume that it is odd under velocity reversal $F_f(-p) = -F_f(p)$.

Considering the joint probability density function $\rho(q,p,t)$ of density of particles in phase space, we have the Kramers equation,
\begin{align}
\pder{\rho}{t} + \frac p m \pder{\rho}{q} -U'(q) \pder{\rho}{p} & = 
\npder{}{p}{2}\left[D(p) \rho\right]
-\pder{}{p}\left[  F(p)  \rho\right]
\label{kra}
\end{align}
where the right-hand side contains the Newtonian streaming terms. A typical example of deterministic force comes from a harmonic potential, $U(q) = \frac m 2 (\omega q)^2$. This potential is even, $U(q) = U(-q)$, and confining ($\lim_{q\to\pm\infty}U(q) = +\infty$). We proceed by considering only potentials sharing these two properties.
These conditions are necessary for the existence of a steady state regime with distribution $\rho_s(q,p)$ that fulfills $\zpder{\rho_s}{t} = 0$. We assume this distribution to exist and to be normalizable. Its averages are denoted by $\mean{\ldots}$.

We now multiply (\ref{kra}) by $q p$ and integrate by parts over $q$ and $p$.
The term including $\partial^2_p [ D(p) \rho(p)]$ provides a null contribution because the associated boundary terms are assumed to vanish. Hence, we get
\begin{align}
\mean{\frac{p^2}m}  - \mean{q U'(q)} = - \mean{q F(p)} .
\label{mve}
\end{align}
This expression represents a generalized virial relation between kinetic energy, a virial term $-\mean{q U'(q)} $, and a nonlinear dissipative term $- \mean{q F(p)}$.
For example, assuming that $F(p)$ reduces to a linear friction force $F_f(p) = -\gamma p/m$ we get
\begin{align}
\mean{\frac{p^2}m} - \mean{q U'(q)} = \gamma \mean{q p}
\end{align}
but here also $\mean{q p}  = \frac m 2 \frac{d}{dt} \mean{q^2} = 0$ in a steady state. Hence we recover the virial theorem
\begin{align}
 \mean{\frac{p^2}m} = \mean{q U'(q)}.
 \label{equi0}
\end{align}
For example, for the harmonic potential it gives
\begin{align}
\mean{\frac{p^2}m} = \mean{m \omega^2 q^2} ,
\label{equi}
\end{align}
namely, the equipartition on the average of kinetic and potential energies.

An additional relation can be obtained multiplying (\ref{kra}) by $p^2$ and integrating by parts over $q$ and $p$.
\begin{align}
0 = \mean{D(p)} + \mean{p F(p)}
\label{kinetic}
\end{align}
Here we have used the fact that $\mean{p U'(q)}=m \frac{d}{dt}\mean{U(q)}$ vanishes in the stationary state.
Note that for systems in contact with a thermal bath, i.e.~such that $F(p)=F_f(p)= -\gamma p/m $ and $D(p)= \gamma k_B T$ (with $k_B$ the Boltzmann constant), \eqref{kinetic} relates the mean kinetic energy to the bath temperature $T$ as $\frac{1}{m} \mean{p^2}_{\text{eq}}=k_B T$. Here $\mean{\dots}_{\text{eq}}$ denotes an average with respect to the equilibrium Gibbs-Boltzmann probability distribution $\rho_{\text{eq}}(p,q) = \frac{1}{Z} e^{-\frac{1}{k_B T}[\frac{p^2}{2 m}+ U(q)]}$, with $Z$ a normalization factor.
 Therefore, in thermal equilibrium the energy equipartition \eqref{equi0} is not only between the particle velocity and position, but also between the particle and the thermal bath degrees of freedom, in the form
\begin{align}
 \mean{\frac{p^2}m}_{\text{eq}} = k_B T=\mean{q U'(q)}_{\text{eq}}.
 \label{equiequi}
\end{align}

However, if $F(p)$ is nonlinear, while (\ref{mve}) remains valid, we may also evaluate other approaches for obtaining simpler relations.
If the dissipative force is polynomial in the momentum, it seems reasonable to plug it directly to (\ref{mve}).
For example, for the case of Rayleigh-Helmholtz friction 
used to describe active Brownian particles \cite{rom12} and molecular motors \cite{bad02}, there is a velocity-dependent force
\begin{align}
F(p) = \alpha p - p^3, \qquad \alpha>0.
\label{acti}
\end{align}
This force corresponds in our notation to a nonlinear friction, which correctly dampens the motion for large velocities ($p>p_0 \equiv\sqrt{\alpha}$) while $F(p)$ is in the same direction of the velocity for $p <  p_0$ and hence it effectively propels the slow particles.
By expanding the last term in the mesoscopic virial equation (\ref{mve}) and collecting also mixed position-velocity terms from $\mean{x F(p)}$, we get
\begin{align}
 \mean{p^2}  - \mean{q U'(q)} = \mean{q p^3}.
\label{mve_acti}
\end{align}
This is simple enough to not need alternative approaches. In the following sections we show that a better option exists for systems displaying non-polynomial forces, such as trapped atoms with Sisyphus cooling, which we now briefly introduce.

\section{Lasers and Sisyphus cooling}

Cold atoms with a Sisyphus cooling mechanism are perfect for illustrating the basic steps of the strategy introduced in the next section.
In the semi-classical approximation, we consider a stochastic motion induced by a laser field on cold atoms \cite{dec15,dec16}:
\begin{align}
  \dot p(t) & =  -U'(q) + \underbrace{ \pder{D(p)}{p}-\frac{\gamma p}{1+(p/p_c)^2}}_{F(p)} + \sqrt{2 D(p)} \cdot \xi(t) \nonumber \\
  \dot q(t) & = p(t) / m
  \label{lange2}
\end{align}
Note that, with respect to the anti-Ito Langevin equation in \cite{dec15,dec16}, the equivalent Ito form \eqref{lange2} acquires an extra term $\zpder{D(p)}{p}$ in the dissipative force $F(p)$. Here, the momentum-dependent diffusion constant has the form
\begin{align}
  D(p) &= D_0 + D_1(p)\nonumber\\
       &= D_0 + \frac{D_1}{1+(p/p_c)^2}
\end{align}
where $p_c$ is the capture momentum above which cooling becomes ineffective,  $D_0$ is the constant part of the diffusion coefficient, and with a slight abuse of notation by $D_1(p)$ we term the momentum-dependent part of the diffusion coefficient and by $D_1$ its amplitude.
Indeed, in the semi-classical approximation, the friction for Sisyphus cooling is
\begin{align}
  F_f(p) = - \dfrac{\gamma p}{1+(p/p_c)^2}
\label{sysi}
\end{align}
This friction has two distinct regimes:
\begin{align}\label{scaling_Ff}
F_f(p) \sim -p & \quad\textrm{for}\quad p\ll p_c \nn
F_f(p) \sim -\frac 1 p & \quad\textrm{for}\quad p\gg p_c 
\end{align}
Note in particular that the friction tends to zero for $p \to \infty$, while it retains a linear character for small momenta.

We will focus on a typical experimental condition in which the confining potential is harmonic,
\begin{align}
U(q) = \frac{m} 2 \omega^2 q^2,
\end{align}
expressed in terms of a typical frequency $\omega/{2 \pi}$ and of the ``mass'' $m$. Given the typical timescale $1/\omega$ and the reference momentum $p_c$, it is convenient to convert the above equations in terms of the dimensionless quantities defined in Table~\ref{tab:1}. We obtain
\begin{align}
  \dot v(t) & =  -U'(x) + \pder{\caD(v)}{v} + F_f(v) + \sqrt{2 \caD(v)} \cdot \xi(t) \nonumber \\
  \dot x(t) & = \Omega v(t)
  \label{lange3}
\end{align}
  In dimensionless units, the harmonic potential becomes $U(x) = \frac 1 2 \Omega x^2$.
  Any Langevin dynamics with generic confining potential $U(x)$ can be converted in its dimensionless version (\ref{lange3}), as long as a typical timescale $1/\omega$ is defined.
  Hence, the force $-U'(x)$ remains written in a generic notation as long as we are not specializing the harmonic case.

\begin{table}[!t]
  \makegapedcells
\begin{center}
\begin{ruledtabular}
\begin{tabular}{ l  l  l }
quantity & & dimensionless \\
\hline
time & $t$ & $t \leftarrow  \gamma t$  \\
momentum & $p$ & $v = \dfrac p {p_c}$ \\
position & $q$ & $x = \dfrac {m \omega}{p_c} q$ \\
angular frequency & $\omega$ & $\Omega = \dfrac{\omega}{\gamma}$ \\
diffusion constants & $D_0$ & $\caD_0 = \dfrac{D_0}{\gamma p_c^2}$ \\
& $D_1$ & $\caD_1 = \dfrac{D_1}{\gamma p_c^2}$ \\
& $D_1(p)$ & $\caD_1(v) = \dfrac{\caD_1}{1+v^2}$ \\
& $D(p)=D_0+D_1(p)$ & $\caD(v) = \caD_0 + \caD_1(v)$ \\
friction force & $F_f(p)$ &  $F_f(v) = - \dfrac{v}{1+v^2} $ \\
harmonic force & $-U'(q) = -\omega^2 q$ & $-U'(x) = -\Omega x$ 
\end{tabular}
\end{ruledtabular}
\end{center}
\caption{With a slight abuse of notation, we use the same letter $t$ both for the physical time and for the dimensionless time. Note that $F_f(v)$ and $\caD(v)$ would be proportional to each other and satisfying a fluctuation-dissipation relation if $D_0=0$.}
\label{tab:1}
\end{table}

\section{Generalized virial relation}

For systems with a peculiar friction term, we show that there is an alternative approach to the straightforward equation (\ref{mve}), which may lead to a simple and elegant generalized equipartition relation involving sums of polynomial terms of the momentum and position, in addition to the usual mean squared momentum and mean squared position. We saw already that the standard virial theorem is found technically by multiplying  the Kramers equation (\ref{kra}) by $qp$ and integrating by parts, then exploiting the fact that the time derivative of $\langle q^2 \rangle$ is zero. We now aim to use a similar strategy, but due to the non-linearities we cannot simply  consider the product $qp$.

In the dimensionless units of Table~\ref{tab:1}, with Kramers equation
\begin{align}
\pder{\rho}{t} + \Omega v \pder{\rho}{x} -U'(x) \pder{\rho}{v} & = 
\npder{}{v}{2}\left[\caD(v) \rho\right]
-\pder{}{v}\left[  F(v)  \rho\right]\,,
\label{kra.d}
\end{align}
the basic idea is to get back a null term of the form $\mean{x v}=0$ after integration over $x,v$ of (\ref{kra.d})
multiplied by $x g(v)$, where $g(v)$ is a function suitable to transform favorably the term $-\pder{}{v}[  F_f(v) \rho]$.

To determine which $g(v)$ makes the trick, consider the integration by parts
\begin{align}
- \intii dx \intii dv\, x\, g(v) \pder{}{v}[  F_f(v) \rho ]
= \nn
= \intii dx \intii dv\, x\, \pder{g(v)}{v}   F_f(v) \rho
\end{align}
It is convenient to choose $g(v)$ such that its derivative $g'(v)\equiv \zpder{g(v)}{v}$ satisfies
\begin{align}
g'(v) F_f(v) = v \qquad \Rightarrow \qquad g'(v)  = \frac{v}{F_f(v)} \,,
\label{trick}
\end{align}
which leads to
\begin{align}
g(v) =  \int dv \frac v {F_f(v)} + \textrm{const}.
\label{gv}
\end{align}
Again, with the usual linear friction $F_f = -v$ this would lead to $g(v) = -v + \text{const}$.
The additive constant is not relevant because it does not enter in (\ref{trick}), thus we could simply choose $g=-v$, as above, to simplify the equations and get to the virial theorem.

In the case of driven cold atoms, by embedding the friction (\ref{sysi}) in (\ref{gv}) we get
\begin{align}
g(v) =  \int dv (1+v^2) + \textrm{const} = \left(v+\frac{v^3}3\right)+\textrm{const}.
\label{gv1}
\end{align}

In general, multiplying the Kramers equation (\ref{kra}) by $x g(v)$ with $g(v)$ obtained through (\ref{gv}), after the usual averaging one gets
\begin{align}
 -\Omega &\mean{v g(v)}+ \mean{g'(v) x\left[U'(x)- \pder{\caD}{v} \right]} =\nonumber \\ &
 = \mean{\caD_0 x g''(v)}  +  \mean{\caD_1(v) x g''(v)}+ \mean{g'(v) x F_f(v)}.
  \label{g2}
\end{align}
where for convenience we have already split the term with $\caD_0$ from the one with fluctuating diffusion coefficient $\caD_1(v)$.
The last average on the right, by construction, transforms to the null term $\mean{x v} =\frac 1 2 \frac{d}{dt} \mean{x^2} = 0$ at steady state. Moreover, if we assume that the velocity-dependent part of the noise intensity satisfies the (non-linear) Einstein relation \cite{klimontovich2012}
\begin{align}
  \caD_1(v) = \alpha \frac{F_f(v)}{v}= \alpha \frac{1}{g'(v)},
  \label{assu}
\end{align}
(with some constant $\alpha$), we obtain that
\begin{align}
  -\caD_1' g' = \caD_1 g''
  \label{assu2}
\end{align}
This is indeed the case for Sisyphus cooling, and generally applies to any system in which the friction $F_f$ and the part of the noise with intensity $\caD_1$ are originated in the same bath.
With the definition (\ref{gv}) and the assumption \eqref{assu},
due to \eqref{assu2} we have that two terms cancel each other in \eqref{g2} and hence it reduces to the main general equation of this work,
\begin{align}
-\Omega \mean{v g(v)} + \mean{g'(v) x U'(x)} & = \caD_0 \mean{x g''(v)}
\label{vir1}
\end{align}
or
\begin{align}
-\mean{v g(v)} + \mean{\frac{v}{F_f(v)} x U'(x)} & = \caD_0 \mean{x \pder{}{v}\left(\frac{v}{F_f(v)}\right)}
\label{vir2a}
\end{align}
Note that the constant in (\ref{gv}) remains irrelevant as long as $\mean{v}=0$. One can check that (\ref{vir2a}) turns into the standard equipartition (\ref{equi}) when $F_f = -v$.

\subsection{Generalized equipartition for Sisyphus cooling}

With Sisyphus friction (\ref{sysi}) with $g(v) = v +v^3/3$ we may simplify (\ref{vir2a}) to
\begin{align}
-\Omega  \mean{v \left(v+\frac{v^3}3\right)} + \mean{(1+v^2) x U'(x)} & = \caD_0 \mean{x \cdot 2 v} = 0
\label{vir2}
\end{align}
in which again we have used $\mean{x v}=0$. Therefore, for cold atoms driven by a laser one should observe
\begin{align}
\Omega \mean{v^2} + \frac \Omega  3 \mean{v^4} = \mean{ x U'(x)} + \mean{ v^2 x U'(x)}.
\label{laser}
\end{align}
With the harmonic potential $U(x) = \frac 1 2 \Omega x^2$ this relation becomes an equation involving simple even moments of velocity and position, 
\begin{align}
\mean{v^2} + \frac 1 3 \mean{v^4} =  \mean{x^2} + \mean{ v^2 x^2} .
\label{laser+h}
\end{align}
This is a generalized form of the equipartition theorem. Arguably, it is more elegant and straightforward than the form one would obtain by applying directly the general equation \eqref{mve}, which would contain a term averaging a ratio $\mean{v x / (1+v^2)}$.

Note that \eqref{laser+h} remains valid even for $p_c \to  \infty$, when the friction is linear and the dynamics is thus in equilibrium.
Indeed, we can check that  $\frac 1 3 \mean{v^4}_{\text{eq}}= \mean{v^2}_{\text{eq}}^2$ thanks to the Gaussian statistics of the velocity distribution, and 
$\Omega \mean{ v^2 x^2}_{\text{eq}}= \mean{ v^2}_{\text{eq}} \mean{ x^2}_{\text{eq}}=\mean{v^2}_{\text{eq}}^2$, because $x$ and $v$ are independent and the equilibrium equipartition \eqref{equi} (in the dimensionless form) holds.

Finally, we mention that other choices for $g(v)$ are possible. Each would yield a different generalized equipartition relation involving higher moments of $x$ and $v$. We provide an example in Appendix~\ref{appe}.

\subsection{Numerical results}

To obtain trajectories for Sisyphus cooling trapped by a harmonic potential, in the simplified setup with $\caD_1=0$, we have integrated numerically \eqref{lange3} by adapting a standard scheme~\cite{siv13} to the case of nonlinear friction force.  Within each time step $dt$, the following algorithm generates the new values of (dimensionless) velocity and position $(v_{t+dt},x_{t+dt})$ from $(v_{t},x_{t})$ and harmonic force $f_t=-\Omega x_t$:
\begin{subequations}
\begin{align}
  c_{t} & = e^{-\frac 1 2 dt/(1+v_t^2)}\\
  v_{t+dt/2} &= c_{t} v_t + \frac{dt}2 f_t + B\, {\cal N}\\
  x_{t+dt}   &= x_t + \Omega\, dt\, v_{t+dt/2}\\
  f_{t+dt}   &= -\Omega x_{t+dt}\\
  c_{t+dt/2} & = e^{-\frac 1 2 dt/(1+v_{t+dt/2}^2)}\\
  v_{t+dt}   &= c_{t+dt/2} v_{t+dt/2} + \frac{dt}2 f_{t+dt} + B\, {\cal N'}
\end{align}
\label{num_int}
\end{subequations}
where $B=\sqrt{(1-e^{-dt})\caD_0}$ is the prefactor of normally distributed random numbers ${\cal N},{\cal N'}$, and where friction terms $c_{t},c_{t+dt/2}$ depend on the velocities $v_t$ and $v_{t+dt/2}$, respectively.

\begin{figure}[!t]
  \includegraphics[width=0.95\columnwidth]{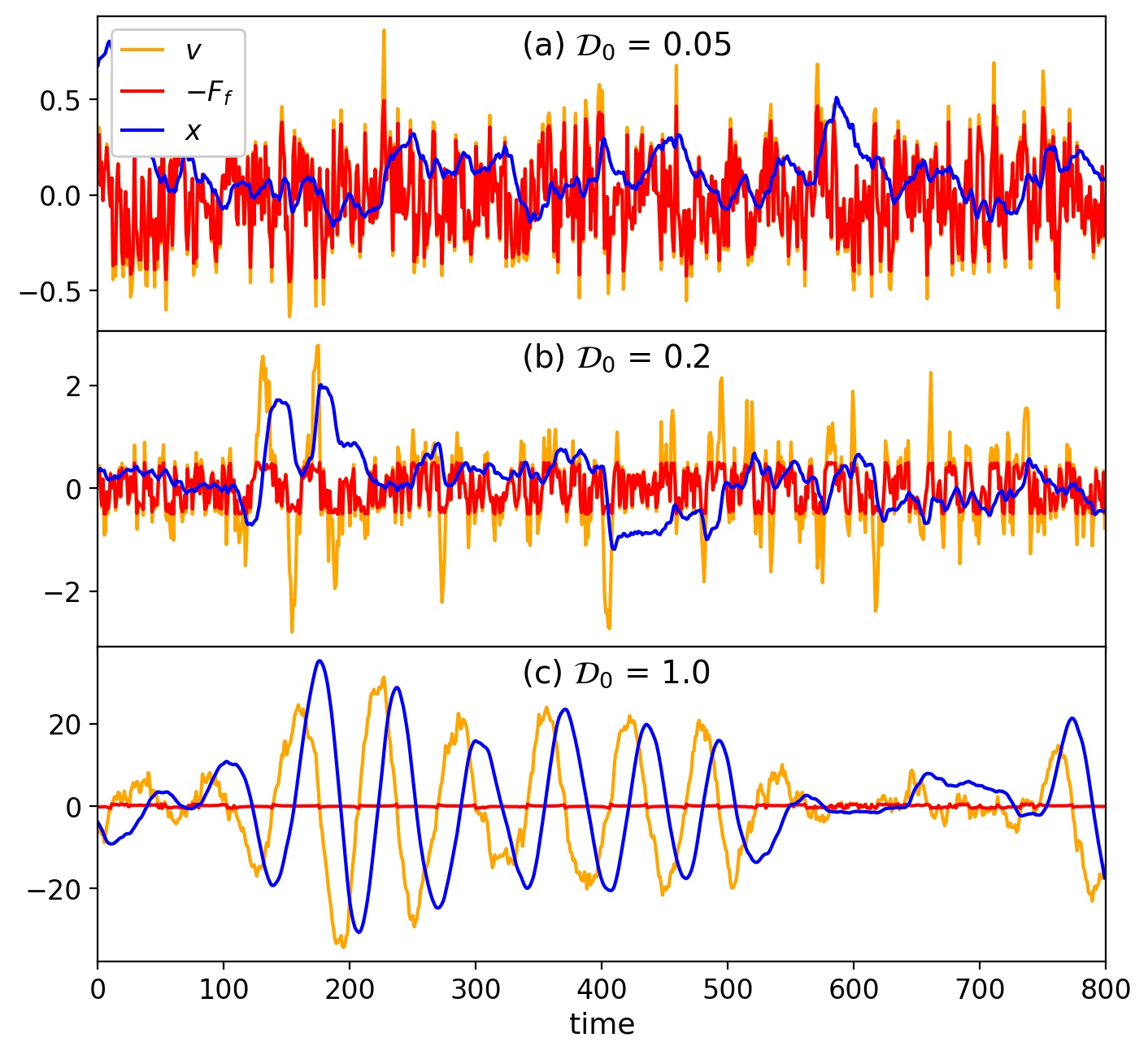}
  \caption{Example of time series of velocity, minus friction force, and position, for a harmonic potential with $\Omega=0.1$, $\caD_1=0$ and (a) $\caD_0=0.05$, (b) $\caD_0=0.2$, (c) $\caD_0=1$. Numerical integrations of \eqref{lange3} follow the scheme \eqref{num_int} with $dt\le 10^{-4}$. }
  \label{fig:ts}
\end{figure}

\begin{figure}[!t]
  \includegraphics[width=0.95\columnwidth]{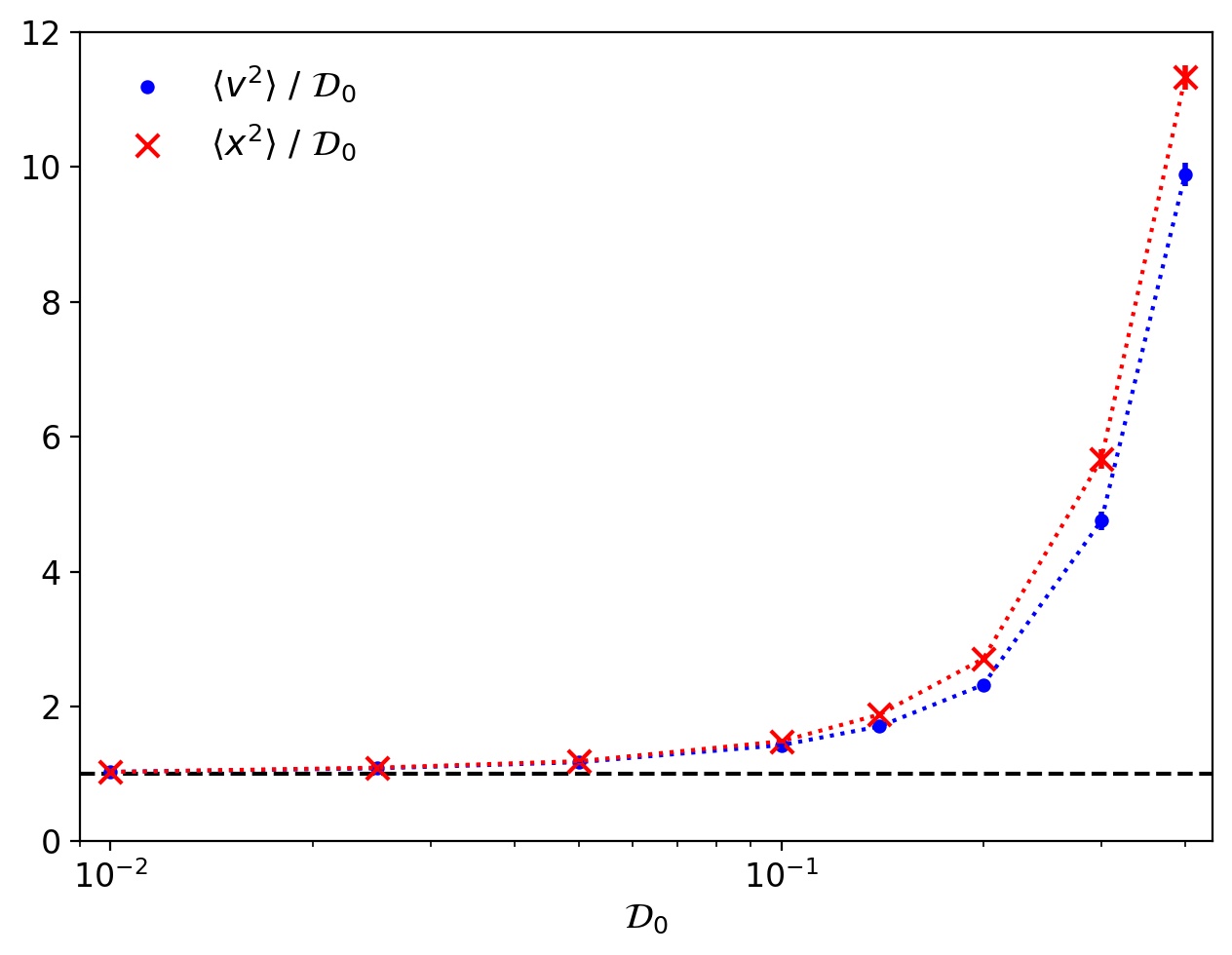}
  \caption{For a harmonic potential with $\Omega=0.1$,  $\mean{v^2}/\caD_0$ and $\mean{x^2}/\caD_0$  as a function of $\caD_0$ ($\caD_1=0$). For $\caD_0 \to 0$ equipartition is approximately valid while it is violated for $0.1 \lesssim \caD_0$. Dotted lines are guides to the eye.}
  \label{fig:xv}
\end{figure}

We use a time step $dt=10^{-4}$ for $\caD_0<0.5$ and $dt=0.5\times 10^{-4}$ for $\caD_0>0.5$. For each set of parameters we collect $N=10^6$ samples, with a sampling time step $\Delta t=1 \ge 10^{4} dt$. The autocorrelation $C_x(i)= \mean{x_j x_{j+i}}$ of the collected data series $\{x_i\}$, $i\le N$,  is used to compute the autocorrelation ``time''  $i^*$, as the smallest $i^*$ such that $C_x(i^*)/C_x(0)<e^{-1}$. The corresponding number of independent samples $N'=N/i^*$ is used to compute errors for each average quantity $\mean{q}=\sum_{i\le N} q_i / N$ as $[(\mean{q^2} - \mean{q}^2)/N']^{1/2}$.

Some parts of time series generated with this method are shown in Fig.~\ref{fig:ts} for three values of $\caD_0$. The friction force essentially equals (minus) the velocity when the system is affected by a weak noise (small $\caD_0$, as in Fig.~\ref{fig:ts}(a), yields $-F_f \simeq v$). Fig.~\ref{fig:ts}(b) shows longer periods of $|F_f| < |v|$ for larger $\caD_0$'s, which are induced by broader variations of the velocity $v$. Finally, Fig.~\ref{fig:ts}(c) displays the broad fluctuations of $x$ and $v$ in the large $\caD_0$ regime, where their density function acquires fat tails \cite{dec15,dec16}.

\begin{figure}[!t]
  \includegraphics[width=0.95\columnwidth]{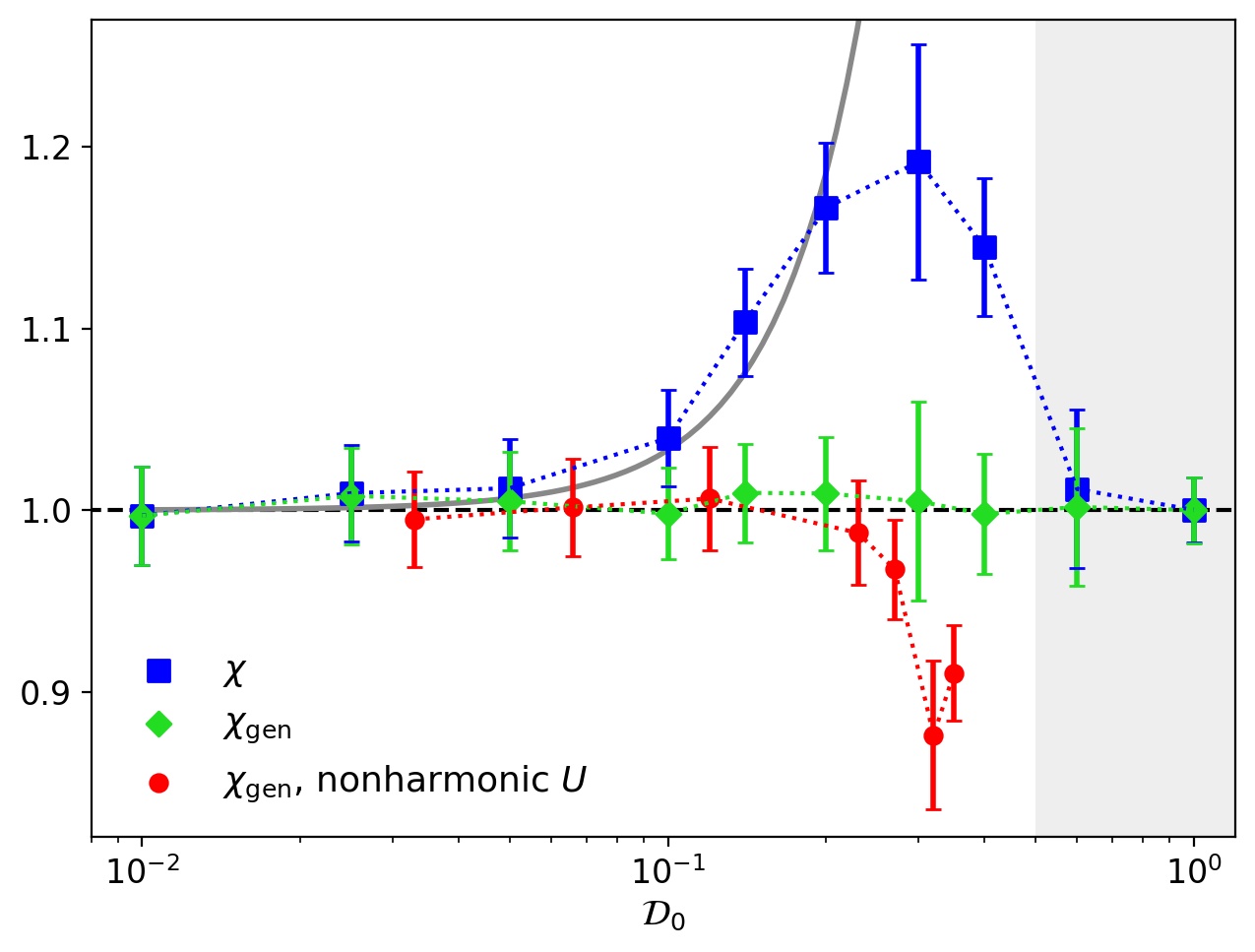}
  \caption{Ratio of equipartition terms vs $\caD_0$ ($\caD_1=0$, $\Omega=0.1$), highlighting the departure from normal equipartition, the validity of the new generalized equipartition (\ref{laser+h}), and the violation of the latter in a nonharmonic system perturbed by a quartic potential energy term. The gray curve is the analytical expansion of $\chi$ up to third order in $\caD_0$, Eq.~26 in Ref.~\cite{dec16}. Dotted lines are guides to the eye and the gray area refers to the regime $\caD_0>1/2$ where sampling is not stationary due to fat tails in the phase space probability density.}
  \label{fig:chi}
\end{figure}

In Fig.~\ref{fig:xv} we plot $\mean{v^2}/\caD_0$ and $\mean{x^2}/\caD_0$ as a function of $\caD_0$, with $\Omega=0.1$. At small values of $\caD_0$ they both tend to $1$, showing that normal equipartition \eqref{equiequi} at ``temperature'' $k_B T \mapsto \caD_0$ is approximately valid in that range.
This is because the friction force becomes linear in $v$ in the limit $\caD_0 \to 0$.
Conversely, standard equipartition breaks down for increasing values of $\caD_0$.
Fig.~\ref{fig:chi} better visualizes the deviation from equipartition by plotting the ratio 
\begin{align}
\chi \equiv \frac{ \mean{x^2} }{\mean{v^2}},
\end{align}
while the ratio
\begin{align}
\chi_{\mathrm{gen}} \equiv \frac{\mean{x^2} + \mean{ v^2 x^2}}{ \mean{v^2} + \frac 1 3 \mean{v^4}}
\end{align}
remains equal to $1$ (within statistical uncertainty) for all values of $\caD_0$, confirming that our new generalized equipartition relation \eqref{laser+h} is valid.

Analytical results valid in the regime of strong confinement ($\Omega \gg 1$) and constant diffusion function ($\caD_1=0$), show that no stationary probability density exists for $\caD_0 > 1$, and some moments, such as the average energy, are time-dependent for $\caD_0 \geq 1/2$ \cite{dec16}.
Notwithstanding, we decide to plot the empirical averages of $v^2$ and $x^2$ up to $\caD_0=1$.
Interestingly, data for $\chi$ suggests that standard yet nonthermal equipartition $\mean{x^2}=\mean{v^2}\ne\caD_0$ may be also satisfied at large values of $\caD_0$.
This finding can be heuristically explained by the fact that for large $\caD_0$, large noise kicks accelerate the particle into a regime of essentially vanishing friction force (see \eqref{scaling_Ff}). The resulting motion is characterized by long periods of quasi-periodic oscillations, with minimal dissipation only at the turning points, see Fig.~\ref{fig:ts}(c). 
Having this picture in mind, we conclude that the standard equipartition may approximately hold even far from equilibrium, as we can guess from \eqref{mve} (still assuming that $\frac{d}{dt}\mean{qp} \simeq 0$) by setting $F(p) \simeq 0$.

Finally, we recall that (\ref{laser+h}) is valid for quadratic degrees of freedom and can be used to detect departures from a purely harmonic potential. By adding a term
$U_\epsilon = \Omega\epsilon x^4/4$ to the confining potential ($\epsilon=10^{-3}$), we see in Fig.~\ref{fig:chi} that indeed $\chi_{\mathrm{gen}}$ departs from $1$ at sufficiently high values of $\caD_0$, where the system can explore the nonharmonic region of the potential. The trend of $\chi_{\mathrm{gen}}$ toward small values is reasonable: with respect to the harmonic case, the more confined motion seems to reduce more the position dependent term $\mean{x^2}+\mean{x^2 v^2}$ than the term  $\mean{v^2}+\mean{v^4/3}$ depending purely on the velocity.

\section{Conclusions}

For the single-particle Langevin equations with nonlinear friction and multiplicative noise we have shown a general method to derive extensions of the virial and equipartition law to nonequilibrium stationary states. In particular, for trapped atoms cooled by the Sisyphus mechanism we have obtained simple explicit expressions that involve only the lowest moment of the atom position and velocity. For the case of harmonic confinement and additive noise, we have numerically verified that our generalized equipartition relation holds for all relevant values of the noise strength, while equilibrium equipartition is broken at intermediate noise strengths. 
These results can be tested in experiments~\cite{afe20}, and
the method introduced here may be extended to other systems displaying non-polynomial friction forces (see e.g. \cite{goy21}).

\section{Acknowledgments}

The support of Israel Science Foundation's grant is acknowledged (EB).

\appendix
\section{Alternate version of generalized equipartition}
\label{appe}
Let us reconsider the more general case $\caD(v) = \caD_0 + \caD_1/(1+v^2)$ in which $\caD_1$ is not null. 
The aim remains to find equations involving polynomial terms of $x$ and $v$. Thus, a suitable $g(v)$ now should remove both the $\sim (1+v^2)^{-1}$ in $\caD(v)$ and a $\sim (1+v^2)^{-2}$ arising from the gradient of $\caD(v)$ in the drift of the stochastic equations.
It turns out that an interesting choice is
\begin{align}
 g(v) &= v + \frac 2 3 v^3 + \frac 1 5 v^5,\\
 g'(v) &= 1 + 2 v^2 + v^4 = (1+v^2)^2, \\
 g''(v) &= 4 v (1+v^2).
\end{align}
Indeed, by using it in \eqref{g2} together with $U' = \Omega x$, we get
\begin{align}
  0  = & \Omega\left[ \mean{v^2} + \frac 2 3 \mean{v^4} + \frac 1 5 \mean{v^6}\right] \\ \nonumber
 & + (4\caD_0-1)\mean{x v^3}
  -  \Omega \left[\mean{x^2} + 2 \mean{x^2 v^2} + \mean{x^2 v^4} \right],
  \label{equiD1}
\end{align}
where we set again $\mean{x v}=0$ thanks to the condition of stationary state.
This is another generalized equipartition relation in which standard deviations of the position and velocity terms are joined by mixed higher order moments.
In (\ref{equiD1}) there is no explicit dependence on the constant $\caD_1$, which enters only implicitly in shaping the steady state averages. Compared to the relation \eqref{laser+h}, this equation is slightly more complex and includes an explicit dependence on the other constant $\caD_0$ entering in the form of the noise strength $\caD(v)$. Note that plugging \eqref{laser+h} into \eqref{equiD1} we can eliminate some higher moments, arriving at
\begin{align}
  \mean{x^2} -\mean{v^2}  + \frac 1 5 \mean{v^6} -\mean{x^2 v^4}
  &= \frac{1-4\caD_0}{\Omega}\mean{x v^3}.
\end{align}
However, these equations involve high moments of velocity and position, hence in practice their precise evaluation requires a better amount of sampling than that needed for (\ref{laser+h}).


%

\end{document}